\documentclass[twocolumn, twocolappendix]{aastex631}
\usepackage{amsmath}
\usepackage{booktabs}
\usepackage{makecell}

\shorttitle{Eccentricity as a Magnifying Glass}
\shortauthors{Salvarese et al.}

\begin{document}

\title{Eccentricity as a Magnifying Glass: Precision Population Inference Enabled by Eccentric Neutron Star--Black Hole Mergers}

\author[0000-0001-9653-1778]{Alberto Salvarese}
\affiliation{Department of Physics, The University of Texas at Austin, Austin, TX 78712, USA}

\author[0000-0001-5403-3762]{Hsin-Yu Chen}
\affiliation{Department of Physics, The University of Texas at Austin, Austin, TX 78712, USA}

\author[0000-0002-7453-6372]{Aaron Zimmerman}
\affiliation{Department of Physics, The University of Texas at Austin, Austin, TX 78712, USA}

\author[0000-0003-4159-8291]{Snehal Tibrewal}
\affiliation{Department of Physics, The University of Texas at Austin, Austin, TX 78712, USA}

\begin{abstract}
The formation history of compact binary systems remains one of the key open questions in astrophysics. 
Theoretical studies generally favor isolated binary evolution for neutron star--black hole (NSBH) systems, which tends to produce nearly circular orbits. 
However, recent analyses of the gravitational-wave event GW200105 indicate that its source has measurable eccentricity, suggesting that alternative formation channels may also contribute. 
It has been shown that the intrinsic parameters of eccentric NSBH mergers, such as the component masses and spins, are much better measured than circular mergers with LIGO--Virgo--KAGRA (LVK) observatories. 
We explore how such eccentricity-enhanced parameter measurements can affect the inference of NSBH formation channels.
We find that sharper measurements of the effective spin parameter $\chi_{\rm eff}$ increase the fraction of systems for which negative values can be confidently identified, 
allowing for the clear measurement of a spin--orbit misaligned event every $\sim 2.5$ eccentric NSBH detections for an isotropically distributed population in the fifth LVK observing run (O5). 
Improved neutron star (NS) mass measurements provide better constraining power for NS mass distributions, potentially revealing structure and tightening the bounds that can be drawn on their upper and lower masses.
Similarly, the recovery of a metallicity-dependent BH mass distribution is improved by eccentricity-enhanced measurements.
Finally, we show that the proposed population-level eccentricity distribution for dynamical-formation channels can be tested by the end of O5.
\end{abstract}


\correspondingauthor{Alberto Salvarese}
\email{alberto.salvarese@utexas.edu}
\email{hsinyu@austin.utexas.edu }
\email{aaron.zimmerman@utexas.edu}
\email{snehaltibrewal@utexas.edu }

\section{Introduction}
A central open question in the study of neutron star--black hole (NSBH) merger formation is whether these systems form predominantly through standard isolated binary evolution, or whether a non-negligible fraction originates from dynamical interactions. 
Gravitational-wave (GW) observations offer a direct way to address this question, because the physical properties of the binaries measured with GW signals reveal information about their formation history.

The detection of orbital eccentricity is one of the clearest signatures of non-isolated formation. In standard isolated binary evolution \citep[e.g.,][]{Dominik_2012, Belczynski_2016, Broekgaarden_2021, Mandel_2022}, GW emission efficiently circularizes the binary during the inspiral \citep[]{peters_1963}, and the resulting systems are therefore expected to enter the sensitive band of ground-based detectors with negligible eccentricity. By contrast, binaries assembled through dynamical interactions \citep[e.g.,][]{Sigurdsson1993, Kulkarni1993, Portegies_Zwart_2000, Mapelli_2021} or in hierarchical multiple systems can retain measurable residual eccentricity when they become detectable \citep[e.g.,][]{Antonini_2017, Samsing_2018, Dall_Amico_2024, Stegmann_2025, Dhurkunde_2025}. Eccentricity can therefore serve as a tracer of the astrophysical origin of compact binary mergers \citep[]{morras2026impacteccentricitypopulationproperties}.

The relevance of eccentricity for NSBH systems has been highlighted by recent analyses of \texttt{GW200105}~\citep{Abbott_2021}, which reported support for nonzero eccentricity \citep[]{Morras:2025xfu, Kacanja_2025, Tiwari_2025, jan2026gw200105detailedstudyeccentricity, Romero_Shaw_2026, Phukon_2026}. This points to the possibility that at least some NSBH mergers form through channels beyond standard isolated binary evolution. 

In addition, eccentricity affects the information that can be extracted from the GW signal. By introducing higher harmonics and modifying the time-frequency evolution of the waveform, eccentricity can help break parameter degeneracies, increase the signal-to-noise ratio (SNR), and reduce uncertainties on intrinsic source parameters \citep[][]{Mikoczi:2012qy, Gondan:2017hbp, Gondan:2018khr}.
In particular, \citet{Tibrewal_inprep} found that GW200105-like eccentric NSBHs can yield significantly more precise measurements of intrinsic parameters, such as component masses and spin, in LIGO--Virgo--KAGRA (LVK) observatories \citep[]{Advanced_LIGO, Advanced_Virgo, KAGRA}. 
Such improvements propagate directly to population analyses, whose constraining power depends on the precision of event-level posteriors \citep[]{Zeeshan_2024}, further revealing the formation histories of these systems.

More precise constraints on the effective spin, $\chi_{\rm eff}$, make it easier to identify negative values, which are produced when the spin orientations lie below the orbital plane. Such measurements would be difficult to reconcile with standard isolated binary evolution, and would instead support formation channels capable of producing spin--orbit misalignment, including dynamical assembly \citep[e.g.,][]{Rodriguez_2016, Stevenson_2017, Farr_2017, PhysRevD.98.084036, Stegmann_2025}. 
Furthermore, improved measurements of the component masses could help distinguish between different neutron star (NS) and black hole (BH) populations and better constrain the properties of their mass distributions \citep[]{Farah_2022, Biscoveanu_2022, Ye_2022, Zhu_2022}. 
Similarly, if eccentricity can be measured across a population, NSBH observations may begin to test population-level predictions for the eccentricity distribution itself \citep[e.g.,][]{Wen_2003, O_Leary_2009, Antonini_2012, Samsing_2014, Antonini_2017, Samsing_2018, Samsing_2025}, including the proposed universal form for dynamically formed mergers \citep[]{rozner2026universaleccentricitydistributiondynamical}. 

In this work, we explore several examples of how eccentricity-enhanced parameter-estimation (PE) affects NSBH population inference and its implications for revealing eccentric NSBH formation histories. 
We construct simulated NSBH populations and generate mock PE samples under both eccentric and quasi-circular orbit assumptions. 
We then use these catalogs to quantify how improved measurements of masses, spins, and eccentricities can strengthen population-level tests of NSBH formation scenarios.

\section{Simulations and methods}
The population distribution of eccentric NSBHs is unknown. We start with a simulated NSBH population in which the NS and BH masses are drawn independently, without imposing a pairing function, and we neglect correlations between component masses and $\chi_{\rm eff}$. For each simulated binary, we draw the luminosity distance from a distribution uniform in comoving volume, assuming a flat-$\Lambda$CDM cosmology with $H_0=67.74\,\mathrm{km}\,\mathrm{s}^{-1}\,\mathrm{Mpc}^{-1}$ and $\Omega_m=0.3075$ \citep[]{Planck_2015}.

For the NS mass distribution, we adopt a fiducial \texttt{Double-peaked} model consisting of two truncated Gaussian components plus a power-law background, which reduces to a uniform component for the fiducial injected value of the power-law index (blue solid line in Figure~\ref{fig:populations}, top panel; see Appendix~\ref{app:simulations} for more details). This choice is motivated by the observed Galactic NS mass distribution and by previous population studies suggesting a bimodal structure, with peaks near $1.3\,M_\odot$ and $1.8\,M_\odot$ \citep[e.g.,][]{Kiziltan_2013, Ozel_2012, antoniadis2016millisecondpulsarmassdistribution, Shao_2020}. We also consider a \texttt{Uniform} NS mass distribution (orange dashed line in Figure~\ref{fig:populations}, top panel) as a test model for the model-comparison analysis.

\begin{figure}
    \centering
    \includegraphics[width=1\columnwidth]{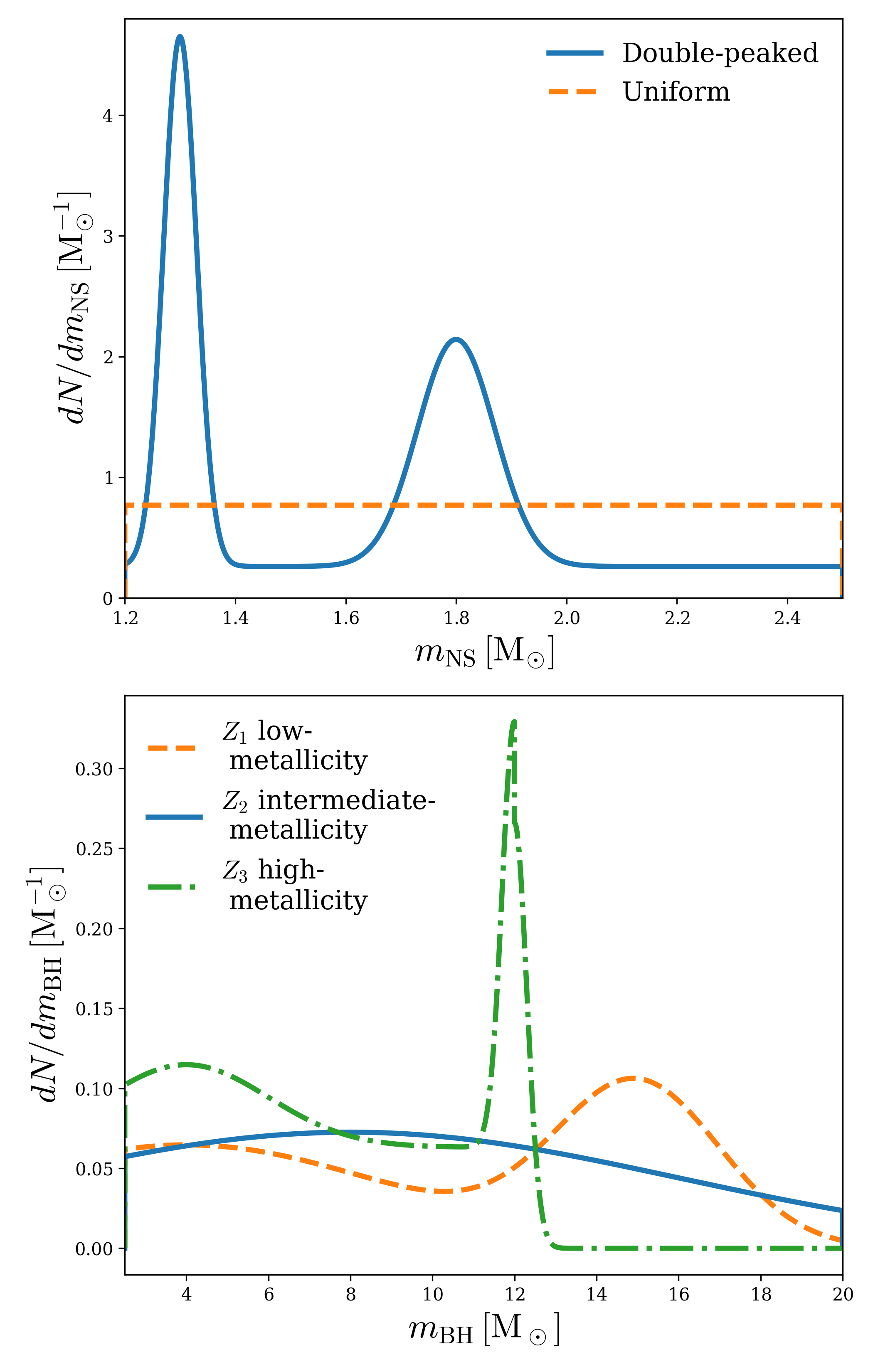}
    \caption{
    Top panel: \texttt{Double-peaked} NS mass distribution (blue solid line; injection) and \texttt{Uniform} NS mass distribution (orange dashed line; test model).
    Bottom panel: BH mass distributions for the three progenitor-metallicity models: $Z_1$ (orange dashed line; test model), $Z_2$ (blue solid line; injection), and $Z_3$ (green dash-dotted line; test model).
    }
    \label{fig:populations}
\end{figure}

For the BH mass distribution, we follow the population-synthesis models considered by \citet{Broekgaarden_2021} and consider three representative distributions corresponding to different progenitor metallicity regimes (see Appendix~\ref{app:simulations}): 
a low-metallicity model $Z_1$ where the mass distribution is described by a mixture of two truncated Gaussian components, an intermediate-metallicity model $Z_2$ described by a single truncated Gaussian, and a high-metallicity model $Z_3$ described by a mixture of two truncated Gaussian components plus a Uniform background. 
We generate the BH mass distribution following the intermediate-metallicity model, and use the other two for our test below. 
We show the corresponding mass distributions in the bottom panel of Figure~\ref{fig:populations}.

We consider a population of dynamically assembled binaries, in which the BH spin tilt is isotropically distributed, and the NS spin is negligible. 
The effective spin is then approximated as
\begin{equation}
    \chi_{\rm eff} \simeq \frac{m_{\rm BH}}{m_{\rm NS}+m_{\rm BH}}\chi_{\rm BH}\cos\theta_{\rm BH},
    \label{eq:chieff_toy}
\end{equation}
with $\cos{\theta_{\rm BH}}\sim \mathcal{U}(-1,1)$ and $\chi_{\rm BH}\sim\mathcal{U}(0,1)$.
 
Motivated by recent work suggesting that dynamically assembled compact-binary mergers may approach a universal eccentricity distribution in the high-eccentricity regime \citep[]{rozner2026universaleccentricitydistributiondynamical}, we draw the true eccentricities over the interval $0.1 \le e \le 0.9$ from a fiducial distribution with scaling
\begin{equation}
    p(\log e) \propto e^{-12/19}.
\end{equation}

Having specified the source intrinsic parameters, we next check which systems are detected and generate mock PE samples.
To determine which simulated binaries pass the detection threshold, 
we add the effects of Gaussian noise to the calculated optimal SNR of the injected binaries, accounting for sky-averaged detector antenna patterns (Appendix \ref{app:simulations}).
For this detection and selection-effects calculation, we assume an A+ design sensitivity representative of the fifth LVK observing run (O5) for a single LIGO detector \citep[\texttt{AplusDesign.txt} in][]{prospects_observing_runs}, and compute the SNR using a minimum frequency of $f_{\min}=20\,\mathrm{Hz}$.
We neglect any explicit dependence of the SNR on orbital eccentricity. Since eccentricity is drawn independently of the intrinsic source parameters entering the selection calculation, this approximation is consistent with our population model and does not introduce an additional selection bias.
We consider an event to be detected if its SNR is at least 8.

We generate synthetic PE samples following the \textsc{GWMockCat} prescription \citep[][]{Farah_2023} for the luminosity distance and component masses, with measurement uncertainties rescaled according to the event SNR. We then extend this mock-PE procedure by generating synthetic posterior samples for $\chi_{\rm eff}$ and eccentricity (see Appendix~\ref{app:simulations} for more details).

We model the effect of eccentricity by reducing the measurement uncertainties relative to the quasi-circular case. This choice is motivated by \citet{Tibrewal_inprep}, who showed that orbital eccentricities in the range $e \sim 0.1$--$0.25$ can reduce the uncertainty in the component masses and $\chi_{\rm eff}$ by factors of ${\sim}5$--$13$ for GW200105-like systems. We therefore adopt a reduction that represents an intermediate choice within that range. We consider two scenarios:

\begin{itemize}
    \item \texttt{Circular}: baseline uncertainties representative of quasi-circular binaries; at SNR 20 for the symmetric mass ratio $\eta$ and effective spin $\chi_{\rm eff}$ this corresponds to $0.021$, and $0.097$, respectively \citep[]{Tibrewal_inprep};
    \item \texttt{Eccentric}: a factor of $10$ reduction in the $\eta$ and $\chi_{\rm eff}$ uncertainties.
\end{itemize}

To reduce statistical fluctuations associated with the random draw of detected events, we repeat each analysis using $30$ independently drawn sub-catalogs with $N_{\rm obs} \in \{5, 10, 15, 20, 30\}$ uniformly sampled from the detected population. 

For the analysis related to NSBH population inference we infer the population properties of our simulated events using hierarchical Bayesian analysis.
For a population model $\mathrm{M}_i$ with hyperparameters $\vec{\Lambda}$, we calculate the Bayesian evidence as
\begin{equation}
    Z_i \equiv p(\mathcal{D}\,|\,\mathrm{M}_i)
    =
    \int d\vec{\Lambda}\,
    \mathcal{L}(\mathcal{D}\,|\,\vec{\Lambda},\mathrm{M}_i)\,
    \pi(\vec{\Lambda}\,|\,\mathrm{M}_i),
    \label{eq:evidence}
\end{equation}
where $\mathcal{L}(\mathcal{D}\,|\,\vec{\Lambda},\mathrm{M}_i)$ is the hierarchical likelihood, which describes the probability of the observed catalog $\mathcal{D}$ given the hyperparameters $\vec{\Lambda}$ of model $\mathrm{M}_i$, and $\pi(\vec{\Lambda}\,|\,\mathrm{M}_i)$ is the corresponding hyperparameter prior (see Appendix~\ref{app:inference} for further details on the adopted formalism). 
We compute the evidences with \texttt{dynesty}, a Python package that implements the nested sampling algorithm \citep[e.g.,][]{skilling, Mukherjee_2006}.

\section{Negative effective spin}
\label{sec:spin}

We start by estimating how improved measurements of the effective spin parameter,
$\chi_{\rm eff}$, could help identify NSBH systems with spin--orbit misalignment. 
A clearly measured negative value of
$\chi_{\rm eff}$ would provide evidence for spin--orbit misalignment, which is
consistent with dynamical assembly.

We compute the fraction of systems expected to have 
$\chi_{\rm eff} < -2\sigma_{\chi_{\rm eff}}$ for both the \texttt{Circular} and \texttt{Eccentric} scenarios. This estimate provides the fraction of  systems for which $\chi_{\rm eff}=0$ would be disfavored at approximately the $2\sigma$ level, offering clear evidence of dynamical assembly.

In the \texttt{Circular} scenario,
we find only $\sim 22\%$ of the detections could offer the clear evidence. 
By contrast, in the \texttt{Eccentric} scenario the much smaller uncertainty on $\chi_{\rm eff}$ increases the fraction to $\sim 43\%$, 
which corresponds to roughly one event every $\sim 2.5$ eccentric NSBH detections.

\section{NS mass distribution}
\label{sec:inference}
Next, we test whether improved NS mass measurements can reveal structure in the NS mass distribution at the population level. 
Specifically, we compare the \texttt{Double-peaked} model used for the simulations to
a simpler \texttt{Uniform} model with hyperparameters $\vec{\Lambda}=\{m^{\rm min},m^{\rm max}\}$. 
The \texttt{Uniform} model is adopted as a minimal reference model against which to test the evidence for structure in the NS mass spectrum and is currently supported by LVK inferences \citep[]{popO3, popO4, pop_gwtc5}.

We quantify the relative support for the two NS population models through the logarithm of the Bayes factor \citep[e.g.,][]{Jeffreys1939-JEFTOP-5, MOREY20166}
\begin{equation}
    \log_{10}\mathcal{B}_{\rm DP,U}
    =
    \log_{10} Z_{\rm DP}
    -
    \log_{10} Z_{\rm U},
    \label{eq:BF}
\end{equation}
so that positive values indicate preference for the \texttt{Double-peaked} model over the \texttt{Uniform} alternative. Since the Bayes factor automatically incorporates the Occam penalty associated with a larger hyperparameter space, any preference for the \texttt{Double-peaked} model reflects support for genuine structure in the NS mass distribution.

\begin{figure}[tb!]
    \centering
    \includegraphics[width=1\columnwidth]{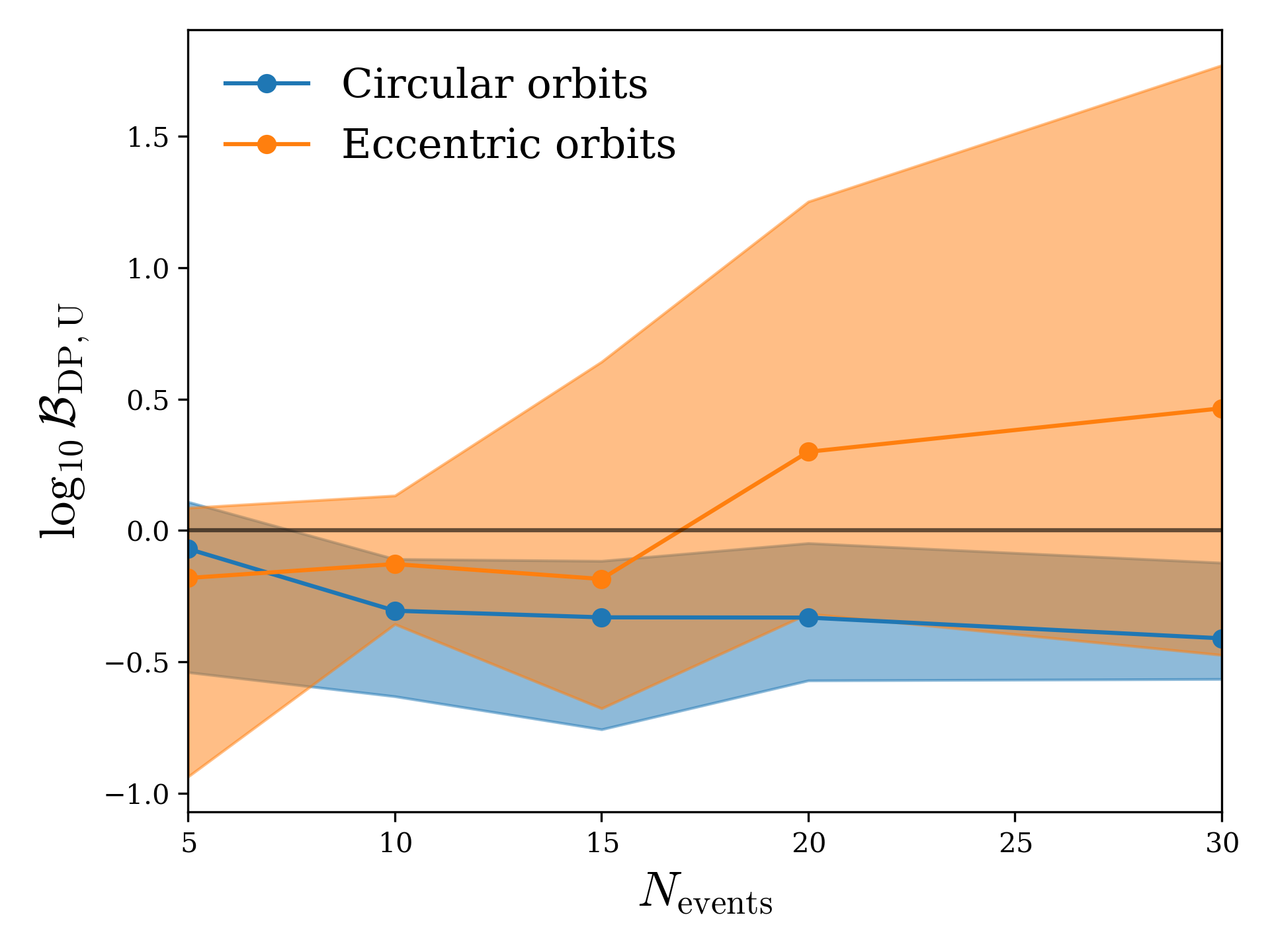}
    \caption{$\log_{10}$ Bayes factor comparison between the fiducial \texttt{Double-peaked} NS mass model and the simpler \texttt{Uniform} NS mass model. We show the median and symmetric 68\% credible interval of $\log_{10}\mathcal{B}_{\rm DP,U}$ over 30 catalog realizations, for the \texttt{Circular} (blue) and \texttt{Eccentric} (orange) PE scenarios. Values greater than zero indicate support for the fiducial \texttt{Double-peaked} model.}
    \label{fig:bayes_comparison}
\end{figure}

Figure~\ref{fig:bayes_comparison} shows $\log_{10}\mathcal{B}_{\rm DP,U}$ as a function of the number of detections, where we plot the median and symmetric 68\% credible interval over 30 independent realizations. In the \texttt{Circular} scenario, the fiducial \texttt{Double-peaked} model is not significantly preferred over the simpler \texttt{Uniform} model even with 30 detections: the larger NS-mass uncertainties obscure the structure of the underlying distribution, while the additional flexibility of the \texttt{Double-peaked} model is penalized by its larger hyperparameter space. By contrast, in the \texttt{Eccentric} scenario, the Bayes factor increases with the number of detections, indicating growing support for the fiducial model. Although the 68\% interval still extends below $\log_{10}\mathcal{B}_{\rm DP,U}=0$, so that some catalog realizations prefer the simpler uniform distribution, this should be compared to the Circular scenario where the entire 68\% interval lies below zero. With 30 detections, the median Bayes factor in favor of the \texttt{Double-peaked} model is larger by up to a factor of $\sim7.5$ relative to the \texttt{Circular} scenario, highlighting the much stronger mass-distribution model-selection power with eccentricity-enhanced PE.

\begin{figure*}[tb]
    \centering
    \includegraphics[width=1\linewidth]{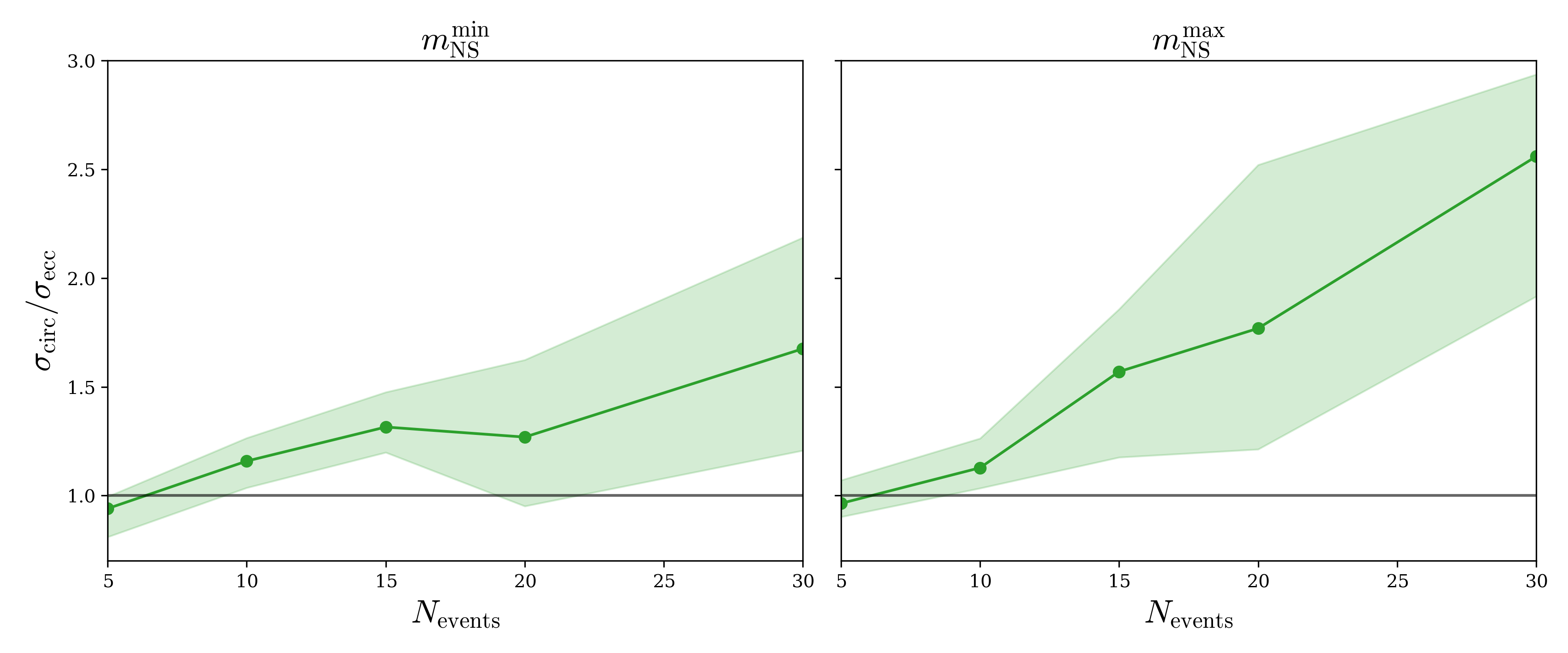}
    \caption{Ratio between the precision of the inferred NS mass hyperparameters obtained under the \texttt{Circular} and \texttt{Eccentric} scenarios. We show the median and symmetric 68\% credible interval of $\sigma_{\rm circ}/\sigma_{\rm ecc}$ over 30 catalog realizations.
    Values greater than unity (black horizontal line) indicate tighter constraints in the \texttt{Eccentric} scenario.
    The left and right panels show the results for $m^{\rm min}_{\rm NS}$ and $m^{\rm max}_{\rm NS}$, respectively.}
    \label{fig:hyperparams}
\end{figure*}
The broader spread in the \texttt{Eccentric} scenario reflects the stronger dependence of the Bayes factor on the specific catalog realization. In the \texttt{Circular} case, the NS-mass uncertainties are sufficiently large that differences between independent realizations are largely washed out. Although each catalog contains a different set of underlying masses, the broad event-level mass estimates make the resulting population inferences relatively similar, leading to comparable Bayes factors across repetitions.
In the \texttt{Eccentric} case, the much smaller mass uncertainties preserve the differences between individual catalog realizations. The inferred Bayes factor therefore depends more strongly on the particular set of masses drawn in each repetition, producing a larger realization-to-realization variation and, consequently, a broader spread in the Bayes factor.

We next investigate how the eccentricity-enhanced measurements of the NS mass propagate into constraints on the NS mass population hyperparameters. 
In Figure~\ref{fig:hyperparams}, we show the ratio between the precision of the inferred hyperparameters obtained under the \texttt{Circular} and \texttt{Eccentric} scenarios as a function of the number of detections (median and symmetric 68\% credible interval over the 30 realizations). Values greater than unity indicate that the hyperparameters are more tightly constrained in the \texttt{Eccentric} scenario. We focus in particular on the minimum NS mass, $m^{\rm min}_{\rm NS}$ (left panel), and the maximum NS mass, $m^{\rm max}_{\rm NS}$ (right panel). The remaining hyperparameters show either no significant change or a comparable level of improvement over the number of detections. We find that the \texttt{Eccentric} scenario yields constraints that are tighter by a factor of $\sim2$ for $m^{\rm min}_{\rm NS}$ and $\sim2.5$ for $m^{\rm max}_{\rm NS}$ with 30 detections, making eccentric NSBH mergers a more informative population for probing NS physics. For example, $m^{\rm max}_{\rm NS}$ provides key observational constraints on the properties of dense nuclear matter \citep[e.g.,][]{Ozel_2012, Lattimer_2016, Chatziioannou_2025}.

\begin{figure*}[tb]
    \centering
    \includegraphics[width=1\linewidth]{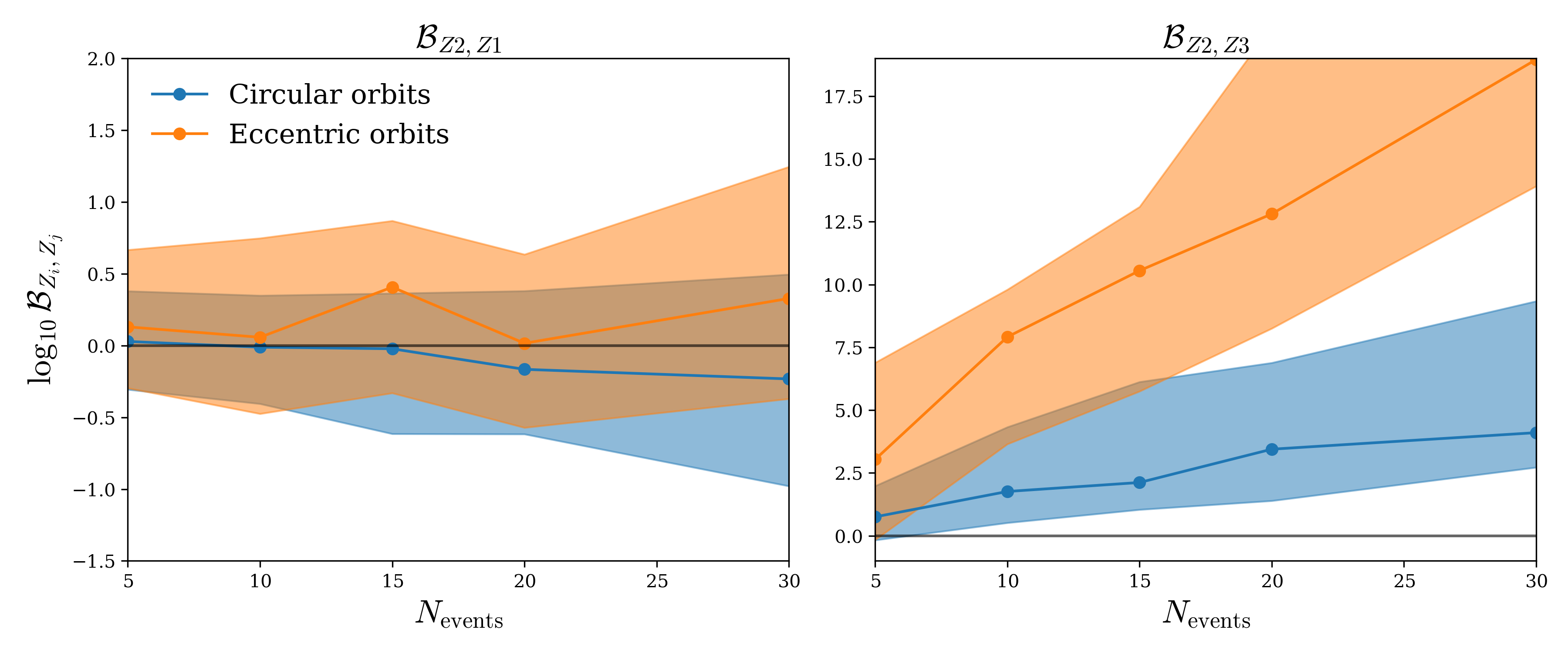}
    \caption{$\log_{10}$ Bayes factors comparing the fiducial intermediate-metallicity BH progenitor model, $Z_2$, against the low- and high-metallicity models, $Z_1$ and $Z_3$. We show the median and symmetric 68\% credible interval over 30 catalog realizations for $\log_{10}\mathcal{B}_{Z_2,Z_1}$ (left panel) and $\log_{10}\mathcal{B}_{Z_2,Z_3}$ (right panel), assuming \texttt{Circular} (blue) and \texttt{Eccentric} (orange). Values greater than zero indicate support for the fiducial intermediate-metallicity model $Z_2$.}
    \label{fig:metallicity}
\end{figure*}

\section{Birth metallicities}
We next investigate whether improved PE for eccentric NSBH systems enhances our ability to distinguish populations arising from different progenitor metallicities. We take the intermediate-metallicity model, $Z_2$, as the fiducial population used to generate the simulated catalogs, and compare it against two alternatives: a low-metallicity model, $Z_1$, and a high-metallicity model, $Z_3$. 

For each simulated catalog, we evaluate the Bayesian evidence under each of the three metallicity models and compare them using the logarithmic Bayes factors
\begin{equation}
    \log_{10}\mathcal{B}_{Z_2,Z_i} = \log_{10}Z_{Z_2}-\log_{10}Z_{Z_i},
\end{equation}
with $i=1,3$. Since our goal is not to infer the hyperparameters of each model, but rather to assess whether the different BH mass-distribution shapes associated with different progenitor metallicities can be distinguished, we impose priors that preserve the overall structure of the corresponding fiducial mass distributions (Appendix~\ref{app:priors}). As in the NS-mass case, we repeat the analysis over $30$ independently drawn sub-catalogs for each choice of $N_{\rm obs}$ and report the results in Figure~\ref{fig:metallicity}.
In the \texttt{Eccentric} scenario, we find systematically stronger support for the true intermediate-metallicity model in both comparisons. In particular, for the $Z_2$-$Z_3$ comparison, eccentricity-enhanced PE increases the Bayes factor in favor of $Z_2$ by up to a factor of $\sim 10^{14}$ at $N_{\rm obs}=30$. 
For the $Z_2$-$Z_1$ comparison, the improvement is more modest. At $N_{\rm obs}=30$, the median Bayes factor across catalog realizations is larger by a factor of $\sim 4$ in the eccentric scenario, although there remains a trial-to-trial scatter. This more modest gain reflects the greater similarity between the two mass-distribution shapes.

\begin{figure}[tb]
    \centering
    \includegraphics[width=1\linewidth]{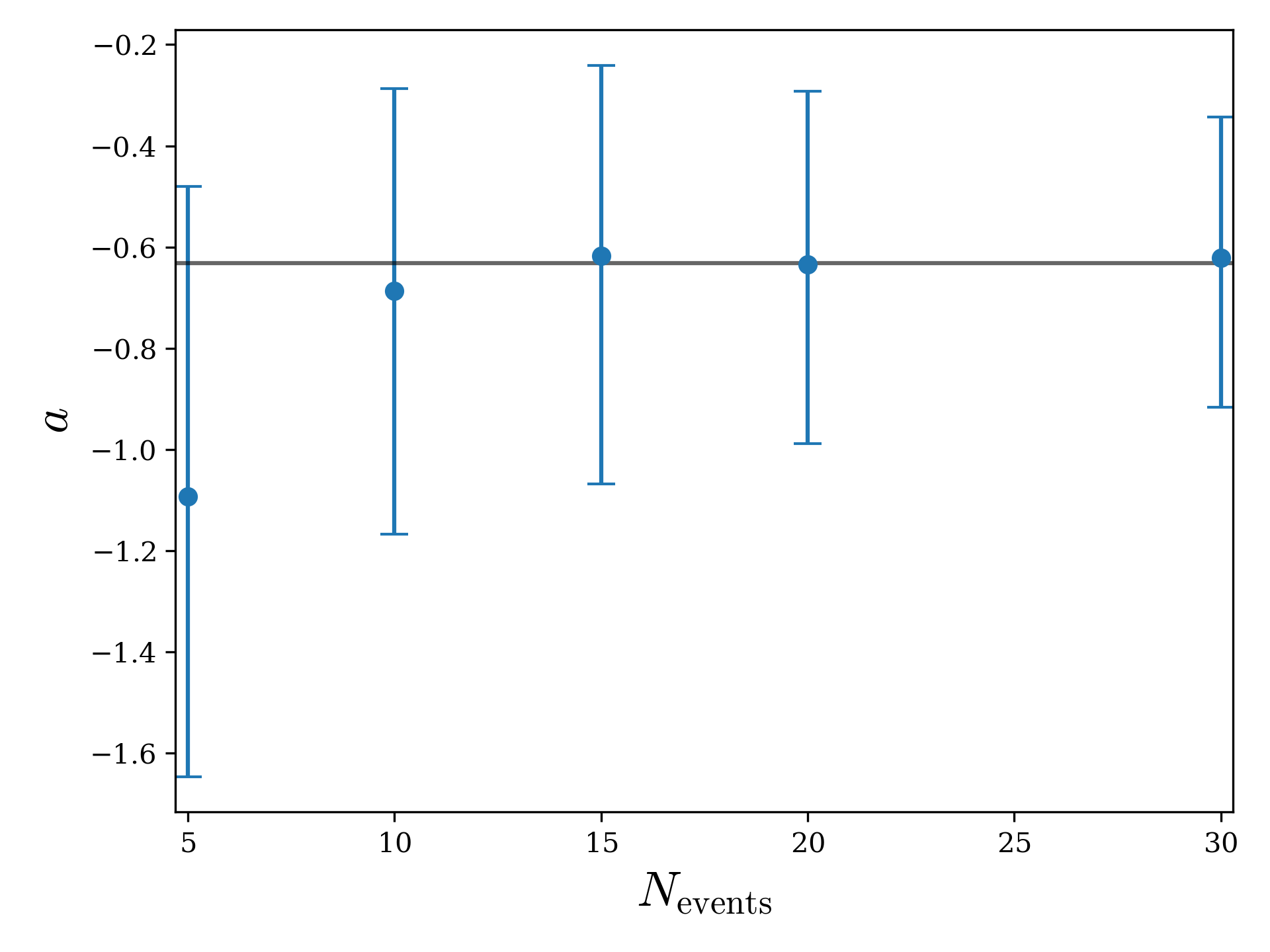}
    \caption{Inferred posterior of the eccentricity-distribution slope $a$. The markers show the median across 30 independent catalog realizations of the posterior median; the error bars show the corresponding median lower and upper bounds of the central 68\% credible intervals. The black horizontal line marks the injected value, $a=-12/19$.}
    \label{fig:CvM}
\end{figure}

\section{Eccentricity distribution}
We investigate how many eccentric NSBH detections are needed to recover the universal eccentricity distribution proposed by \citet{rozner2026universaleccentricitydistributiondynamical}. We use hierarchical Bayesian inference assuming a power-law eccentricity distribution, 
\begin{equation}
p(e) = \frac{a}{e_{\rm max}^a - e_{\rm min}^a} e^{a-1},
\end{equation} 
over the fixed interval $e\in[0.1,0.9]$, and infer only the slope $a$. This choice reflects our goal of testing the recovery of the predicted trend with a finite number of eccentric NSBH detections. 

In Figure~\ref{fig:CvM}, we summarize the inferred posterior on $a$ across 30 independent catalog realizations. For each realization, we compute the posterior median and symmetric 68\% credible interval; the figure shows the median of the posterior medians, with the lower and upper bounds given by the median lower and upper credible limits across the 30 realizations. We find that both the accuracy and precision improve as the number of detections increases, with the inferred posterior converging toward the fiducial trend $a=-12/19$.

\section{Discussion}
Eccentric NSBH mergers are powerful probes of compact-binary formation: eccentricity can point to non-isolated formation, while the associated waveform morphology can sharpen measurements of intrinsic source properties. Although eccentricity measurements can be sensitive to waveform and prior assumptions \citep[]{clarke2026universalframeworkidentifyeccentric}, a sample of confidently identified eccentric systems would provide a valuable opportunity to test formation-related population properties. 

In this work, we quantify how the improved PE expected for eccentric NSBH signals propagates to population-level inference. We focus on four related applications: assessing the formation-channel information contained in sharper $\chi_{\rm eff}$ measurements, resolving structure in the NS mass distribution and tightening its hyperparameter constraints, distinguishing BH progenitor-metallicity models, and testing population-level consistency with a proposed eccentricity distribution. 

For a three-year O5 observing run, rescaling the observed detection rate by the sensitive volume gives $\sim 10$ eccentric NSBH detections. We show that this number of events already enables useful population-level inference in several of the tests considered.
Using our fiducial population, we find that only $\sim 2.5$ eccentric NSBH detections are needed to identify spin--orbit misalignment, providing a clear signature of non-isolated assembly. With 10 eccentric NSBH detections, the correct \texttt{Double-peaked} NS mass distribution can be distinguished from the \texttt{Uniform} model with a factor of 1.5 larger Bayes factor than in the \texttt{Circular} scenario, enabling more informative comparisons with the Galactic NS mass distribution and other possibilities. Similarly, 
the critical population hyperparameters connected to NS nuclear matter properties, $m_{\rm min}^{\rm NS}$ and $m_{\rm max}^{\rm NS}$, can also be measured more precisely, with constraints improving by a factor of $\sim 1.2$.
The correct injected BH progenitor metallicity model can be identified more robustly, providing a potential probe of stellar-evolution processes related to compact binary formation. 
Finally, the proposed population-level eccentricity distribution for dynamical-formation channels can be verified with the $\sim 10$ eccentric NSBH detections.

Across these applications, we show that the constraining power improves as the number of eccentric NSBH detections increases. In the larger catalogs considered in this paper, the constraints on the underlying population become progressively sharper than those obtained for the $\sim 10$ detection case.

We note that our quantitative results depend on the improvement in parameter-estimation precision for eccentric systems. We adopt a factor of 10 reduction in the mass and $\chi_{\rm eff}$ uncertainties as an intermediate choice within the range found in \cite{Tibrewal_inprep}. More conservative reductions could weaken the quantitative gains reported here, while the general trend of improved population inference with sharper intrinsic-parameter measurements is expected to remain. For example, adopting the most conservative factor-of-5 reduction, we find that in the \texttt{Eccentric} scenario the Bayes factor in favor of the \texttt{Double-peaked} model is still $\sim 3$ times larger than in the \texttt{Circular} scenario.

Overall, our results show that eccentric NSBH detections could play a unique role in compact-binary population studies. Since eccentricity both traces non-isolated formation and sharpens measurements of intrinsic source parameters, eccentric NSBH mergers could provide useful insights into the origin and evolution of compact binaries.

\section*{Acknowledgments}
The authors would like to thank Geraint Pratten for the LIGO Scientific Collaboration internal review, and Marco Dall'Amico and Isobel Romero-Shaw for useful comments and discussions.
A.S. and H.-Y. C. are supported by the National Science Foundation under Grant PHY-2308752 and Department of Energy Grant DE-SC0025296. S.T. is supported by NSF grant PHY-2207780. A.Z. is supported by NSF Grant PHY-2308833. 
The authors are grateful for computational resources provided by the LIGO Laboratory and supported by National Science Foundation grants PHY-0757058 and PHY-0823459. This material is based upon work supported by NSF's LIGO Laboratory which is a major facility fully funded by the National Science Foundation. 
This work has preprint numbers UT-WI-19-2026 and LIGO-P2600330.

\software{ \texttt{numpy} \citep[]{2020NumPy-Array}, \texttt{matplotlib} \citep[]{matplotlib}, \texttt{dynesty} \citep[]{Speagle2019}, \texttt{jax} \citep[]{jax2018github}, \texttt{numpyro} \citep[]{numpyro}}

\appendix

\section{Simulations}
\label{app:simulations}
We adopt a fiducial \texttt{Double-peaked} model for the NS mass distribution. Specifically, we construct a mixture between two truncated Gaussian distributions $\mathcal{TN}(\mu_1,\sigma_1,m^{\rm min}, m^{\rm max})$ and $\mathcal{TN}(\mu_2,\sigma_2,m^{\rm min}, m^{\rm max})$, and one power-law distribution $\mathcal{P}(\alpha, m^{\rm min}, m^{\rm max})$ as
\begin{equation}
    \begin{split}
        p(& m_{\rm NS}|\mu_1, \sigma_1, \mu_2, \sigma_2, m^{\rm min}, m^{\rm max}, \alpha) \\
        &= f_1\mathcal{TN}(\mu_1,\sigma_1,m^{\rm min}, m^{\rm max})\\ 
        &+f_2\mathcal{TN}(\mu_2, \sigma_2,m^{\rm min}, m^{\rm max})\\
        &+(1-f_1-f_2)\mathcal{P}(\alpha,m^{\rm min}, m^{\rm max}),
    \end{split}
\end{equation}
where
\begin{equation}
    \mathcal{P}(\alpha, m_{\min}, m_{\max}) =
    \begin{cases}
        \dfrac{1+\alpha}
        {m_{\max}^{1+\alpha} - m_{\min}^{1+\alpha}} m^{\alpha},
        & \alpha \neq -1, \\[1.2em]
        \dfrac{1}{\ln(m_{\max}/m_{\min})}m^{\alpha},
        & \alpha = -1.
    \end{cases}
\end{equation}

To simulate our fiducial population, we adopt the following hyperparameter values:
\begin{equation}
\begin{split}
    \{\mu_1 &, \sigma_1, \mu_2, \sigma_2, m^{\rm min}, m^{\rm max}, \alpha, f_1, f_2\}\\
    = & \{1.3 \,{\rm M}_\odot,0.03\, {\rm M}_\odot, 1.8\,{\rm M}_\odot, 0.07\,{\rm M}_\odot, \\
    & 1.2\,{\rm M}_\odot, 2.5\,{\rm M}_\odot, 0, 1/3, 1/3\}.
\end{split}
\end{equation}
With $\alpha=0$, the power-law component reduces exactly to a uniform distribution over $[m^{\rm min},m^{\rm max}]$.

For the three BH mass distributions based on the assumed progenitor metallicity, we used the following probability distributions:
\begin{itemize}
    \item $Z_1$ (low metallicity): mixture of two truncated Gaussians,
    \begin{equation}
    \begin{split}
        p(& m_{\rm BH}\,|\,\mu_1,\sigma_1,\mu_2,\sigma_2,m^{\rm min},m^{\rm max})\\
        &= \frac{1}{2}\mathcal{TN}(\mu_1,\sigma_1,m^{\rm min},m^{\rm max})\\
        &+ \frac{1}{2}\mathcal{TN}(\mu_2,\sigma_2,m^{\rm min},m^{\rm max}),
    \end{split}
    \end{equation}
    with fiducial hyperparameters
    \begin{equation}
        \begin{split}\{\mu_1 &,\sigma_1,\mu_2,\sigma_2,m^{\rm min},m^{\rm max}\}\\
        =&
        \{4,\,5,\,15,\,2,\,2.5,\,20\}\,{\rm M}_\odot.
    \end{split}
    \end{equation}

    \item $Z_2$ (intermediate metallicity): single truncated Gaussian,
    \begin{equation}
    \begin{split}
        p(& m_{\rm BH}\,|\,\mu,\sigma,m^{\rm min},m^{\rm max})\\
        &=
        \mathcal{TN}(\mu,\sigma,m^{\rm min},m^{\rm max}),
    \end{split}
    \end{equation}
    with fiducial hyperparameters
    \begin{equation}
        \{\mu,\sigma,m^{\rm min},m^{\rm max}\}
        =
        \{8,\,8,\,2.5,\,20\}\,{\rm M}_\odot.
    \end{equation}

    \item $Z_3$ (high metallicity): mixture of two truncated Gaussians plus a Uniform component,
    \begin{equation}
    \begin{split}
        p(& m_{\rm BH}\,|\,\mu_1,\sigma_1,\mu_2,\sigma_2,m^{\rm min},m^{\rm max},m^{\rm max}_{\mathcal{U}})\\
        &=
        \frac{1}{5}\mathcal{TN}(\mu_1,\sigma_1,m^{\rm min},m^{\rm max})\\
        &+
        \frac{1}{5}\mathcal{TN}(\mu_2,\sigma_2,m^{\rm min},m^{\rm max})\\
        &+
        \frac{3}{5}\,\mathcal{U}(m^{\rm min},m^{\rm max}_{\mathcal{U}}),
    \end{split}
    \end{equation}
    with fiducial hyperparameters
    \begin{equation}
    \begin{split}
        \{\mu_1 &,\sigma_1,\mu_2,\sigma_2,m^{\rm min},m^{\rm max},m^{\rm max}_{\mathcal{U}}\}\\
        =&
        \{4,\,2,\,12,\,0.3,\,2.5,\,20,\,12\}\,{\rm M}_\odot,
    \end{split}
    \end{equation}
\end{itemize}
where each mixture is normalized over the corresponding mass range. 

We generate $N_{\rm sources}=5\times10^4$ NSBH systems and $N_{\rm inj}=10^6$ injections from the fiducial NS and BH mass distributions, namely the \texttt{Double-peaked} NS model and the intermediate-metallicity $Z_2$ BH model. Given the sharp features in the NS mass distribution, this injection set provides adequate coverage of the region of parameter space populated by the simulated sources, reducing potential systematic biases in the estimate of the selection function.

In this work, we neglect any explicit dependence of the SNR on orbital eccentricity. While eccentricity can, in principle, affect the detectability of a source, our simulated NSBH populations do not include correlations between eccentricity and the intrinsic source parameters entering the selection calculation. Therefore, neglecting eccentricity in the injection campaign is consistent with the assumptions of the simulation and does not introduce an additional selection bias.

We follow the \textsc{GWMockCat} prescription to generate mock PE samples for the luminosity distance and component masses, assuming the A$^+$ design sensitivity for a single LIGO detector. In this procedure, mock observed values and PE samples are first generated for the chirp mass $\mathcal{M}_{\rm c}$, symmetric mass ratio $\eta$, luminosity distance $D_L$, SNR $\rho$, and angular projection factor $\Theta$. The angular projection factor accounts for the distribution of source sky locations and inclinations relative to the detector \citep{Finn_1993}. The sampled values of $\mathcal{M}_{\rm c}$ and $\eta$ are then converted to component masses. The uncertainties on $\mathcal{M}_{\rm c}$, $\eta$, $\rho$, and $\Theta$ are specified as user-defined inputs at the detection threshold $\rho_{\rm th}=8$ and rescaled according to the SNR of each event (see \citealt{Farah_2023} and \citealt{Ezquiaga_2021} for additional details). We set the baseline uncertainties to match those reported in \citet{Tibrewal_inprep} (Table~\ref{tab:default_uncertainties}).

\begin{figure*}
    \centering
    \includegraphics[width=0.5\linewidth]{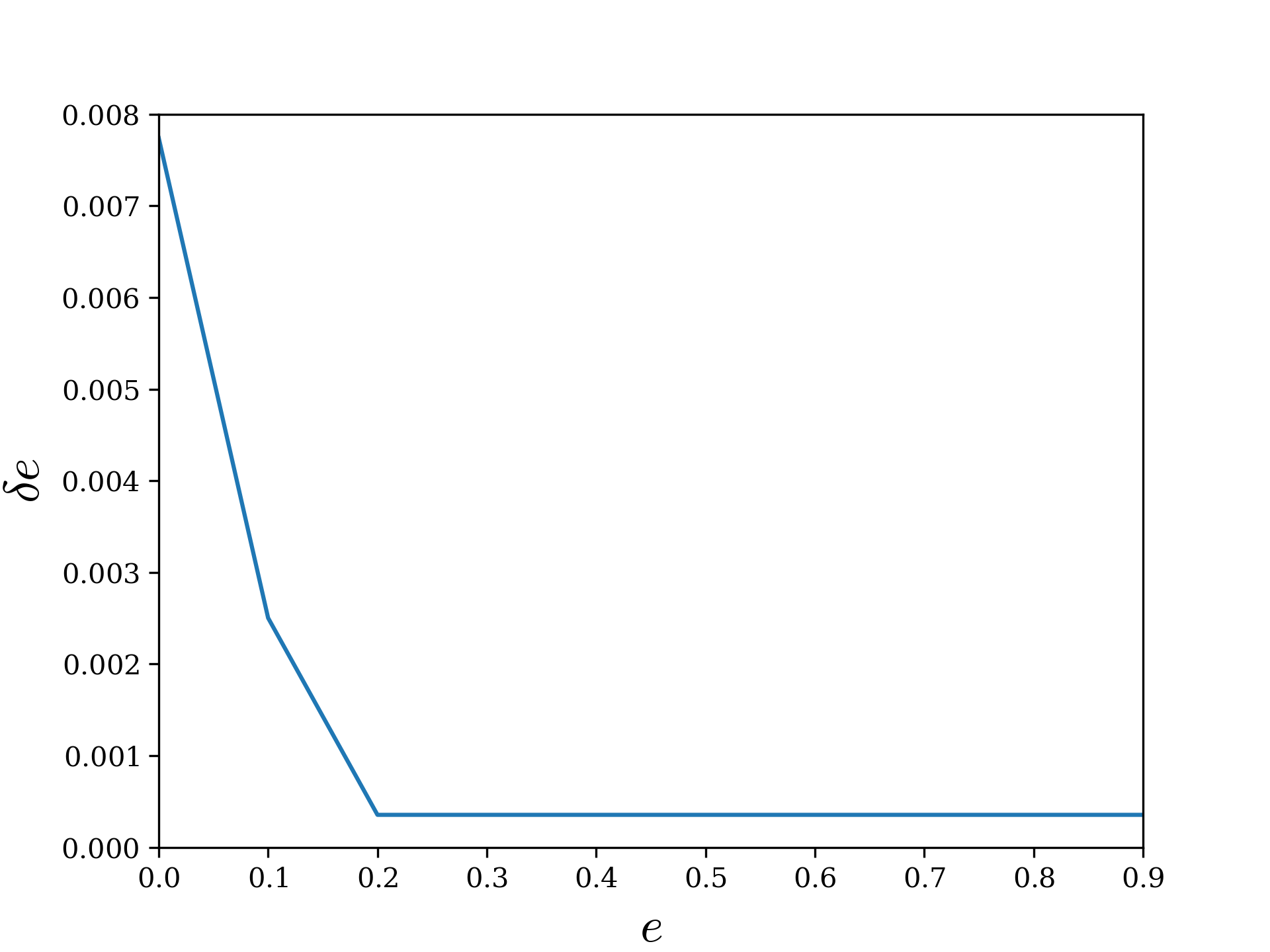}
    \caption{Adopted eccentricity uncertainty $\delta e$ as a function of eccentricity. This uncertainty model is used to generate the eccentricity PE samples adopted in the analysis.}
    \label{fig:eccentricity_uncertainty}
\end{figure*}

For $\chi_{\rm eff}$ and eccentricity, which are not included in the default \textsc{GWMockCat} prescription, we generate PE samples separately. We draw from truncated Gaussian distributions centered on the mock observed values, with widths set by $\delta\chi_{\rm eff}$ and $\delta e$, respectively. For the effective spin, we adopt $\delta\chi_{\rm eff}=0.097292$ in the \texttt{Circular} scenario and $\delta\chi_{\rm eff}=0.0097292$ in the \texttt{Eccentric} scenario. For eccentricity, we determine $\delta e$ by interpolating the uncertainties reported by \citet{Tibrewal_inprep}. For eccentricities larger than the maximum value considered in that study, $e^{\rm max}=0.25$, we fix the uncertainty to the corresponding value, $\delta e(e^{\rm max})$. The resulting uncertainty model is shown in Figure~\ref{fig:eccentricity_uncertainty}. The truncated Gaussian distributions are restricted to the physical intervals $\chi_{\rm eff}\in[-1,1]$ and $e\in[0.1,0.9]$.

\begin{table}[]
    \centering
    \begin{tabular}{cccc}
    \toprule
         $\delta\Theta$ & $\delta\mathcal{M}_c$ & $\delta\eta$  & $\delta\rho$ \\
    \midrule
         $5.0\times 10^{-2}$  & $4.4\times 10^{-3}$ M$_{\odot}$ & $5.0\times 10^{-2}$ & 1 \\
    \bottomrule
    \end{tabular}
    \caption{Default uncertainties adopted in the \textsc{GWMockCat} simulations at the detection threshold $\rho=8$.}
    \label{tab:default_uncertainties}
\end{table}

\section{Inference}
\label{app:inference}
\restartappendixnumbering
Because our goal is to infer the \emph{shape} of the population rather than the absolute merger rate, we use the hierarchical likelihood conditioned on detection
\citep[e.g.,][]{Loredo_selection, Vitale2020}:
\begin{equation}
    \mathcal{L}(\mathcal{D}\,|\,\vec{\Lambda},\mathrm{M})
    \propto
    \prod_{i=1}^{N}
    \frac{
    \int d\vec{\theta}\,
    \mathcal{L}(d_i\,|\,\vec{\theta})\,
    p(\vec{\theta}\,|\,\vec{\Lambda},\mathrm{M})
    }
    {\xi(\vec{\Lambda},\mathrm{M})},
    \label{eq:hierL}
\end{equation}
where $\vec{\theta} = \left\{m_{\rm NS}, m_{\rm BH}, D_{\rm L}\right\}$ denotes the event-level parameters, $\vec{\Lambda}$ is the population hyperparameters of the model $\mathrm{M}$, and
\begin{equation}
    \xi(\vec{\Lambda},\mathrm{M})
    =
    \int d\vec{\theta}\,
    p_{\rm det}(\vec{\theta})\,
    p(\vec{\theta}\,|\,\vec{\Lambda},\mathrm{M})
\end{equation}
is the detection efficiency.

In practice, the event-level integrals are evaluated via Monte Carlo integration over the $N_s$ PE posterior samples $\vec{\theta}_i^{(j)}$ for each event:
\begin{equation}
    \int d\vec{\theta}\,
    \mathcal{L}(d_i\,|\,\vec{\theta})\,
    p(\vec{\theta}\,|\,\vec{\Lambda},\mathrm{M})
    \approx
    \frac{1}{N_s}
    \sum_{j=1}^{N_s}
    \frac{
    p\!\left(\vec{\theta}_i^{(j)}\,|\,\vec{\Lambda},\mathrm{M}\right)
    }{
    \pi\!\left(\vec{\theta}_i^{(j)}\right)
    },
    \label{eq:MC_event}
\end{equation}
where $\pi(\vec{\theta})$ is the prior adopted in the single-event PE analysis.

The selection factor is estimated from the injection campaign as \citep[]{Farr_2019}
\begin{equation}
    \xi(\vec{\Lambda},\mathrm{M})
    \approx
    \frac{1}{N_{\rm inj}}
    \sum_{k=1}^{N_{\rm inj}^{\rm det}}
    \frac{
    p(\vec{\theta}^k\,|\,\vec{\Lambda},\mathrm{M})
    }{
    p_{\rm draw}(\vec{\theta}^k)
    },
    \label{eq:xi}
\end{equation}
where the sum runs over the detected injections and
$p_{\rm draw}(\vec{\theta})$ is the distribution from which injections are generated, i.e., the same model adopted for the population simulation (Appendix \ref{app:simulations}). Because we are not considering any pairing function between BH and NS masses, we can separate the population prior $p(\vec{\theta}|\vec{\Lambda},\mathrm{M})$, and infer separately the NS and BH properties:
\begin{equation}
    \begin{split}
    p(\vec{\theta}^k|\vec{\Lambda},\mathrm{M}) = &\: p(m_{\rm NS}^k|\vec{\Lambda}_{\rm NS},\mathrm{M})p(m_{\rm BH}^k|\vec{\Lambda}_{\rm BH},\mathrm{M})\\
    &\times p(D_{\rm L}^k|H_0, \Omega_m).
    \end{split}
\end{equation}

We also note that since we are only inferring population properties, we can fix the cosmological parameters $H_0$ and $\Omega_m$ to the fiducial injected values.

\section{Priors}
\label{app:priors}
Tables~\ref{tab:NS_priors} and~\ref{tab:BH_priors} report the adopted priors for the NS and BH mass distribution analysis, respectively.

\begin{table*}[]
    \centering
    \small
    \setlength{\tabcolsep}{4pt}
    \begin{tabular}{ccccccc}
    \toprule
    $\mu_1$ 
    & $\sigma_1$ 
    & $\mu_2$ 
    & $\sigma_2$ 
    & $m^{\rm min}$ 
    & $m^{\rm max}$ 
    & $\alpha$ \\
    \midrule
    &&&\texttt{Double-peaked}&&&\\
    \midrule
    $\mathcal{U}(0.7,1.5)$
    & $\mathcal{U}(0.01,0.51)$
    & $\mathcal{U}(\max[1.5,\mu_1],3.6)$
    & $\mathcal{U}(0.01,0.51)$
    & $\mathcal{U}(0.7,\min[1.8,\mu_1])$
    & $\mathcal{U}(\max[1.8,\mu_2],4.8)$
    & $\mathcal{U}(-2,2)$ \\
    \midrule
    &&&\texttt{Uniform}&&&\\
    \midrule
    --
    & --
    & --
    & --
    & $\mathcal{U}(0.7,1.8)$
    & $\mathcal{U}(1.8,4.8)$
    & -- \\
    \bottomrule
    \end{tabular}
    \caption{Priors adopted for the NS mass-distribution models. The \texttt{Double-peaked} model uses conditional bounds that enforce $\mu_1\leq\mu_2$, $m^{\rm min}\leq\mu_1$, and $m^{\rm max}\geq\mu_2$. The \texttt{Uniform} model is defined only by the lower and upper mass bounds. Mass parameters are given in units of ${\rm M}_\odot$.}
    \label{tab:NS_priors}
\end{table*}

\begin{table*}[]
    \begin{tabular}{ccccccc}
    \toprule
    $\mu_1$ 
    & $\sigma_1$ 
    & $\mu_2$ 
    & $\sigma_2$ 
    & $m^{\rm min}$ 
    & $m^{\rm max}$ 
    & $m^{\rm max}_{\mathcal{U}}$ \\
    \midrule
    &&&$Z_1$&&&\\
    \midrule
    $\mathcal{U}(3.5,8.0)$
    & $\mathcal{U}(4.5,5.5)$
    & $\mathcal{U}(14.5,15.5)$
    & $\mathcal{U}(1.5,2.5)$
    & $\mathcal{U}(1.5,3.0)$
    & $\mathcal{U}(17,23)$
    & -- \\
    \midrule
    &&&$Z_2$&&&\\
    \midrule
    $\mathcal{U}(5,11)$
    & $\mathcal{U}(3,13)$
    & --
    & --
    & $\mathcal{U}(0.5,3.5)$
    & $\mathcal{U}(17,23)$
    & -- \\
    \midrule
    &&&$Z_3$&&&\\
    \midrule
    $\mathcal{U}(3,7)$
    & $\mathcal{U}(1,3)$
    & $\mathcal{U}(10,14)$
    & $\mathcal{U}(0.1,0.4)$
    & $\mathcal{U}(1.5,3)$
    & $\mathcal{U}(17,23)$
    & $m^{\rm max}_{\mathcal{U}}\sim\mathcal{U}(10,14)$ \\
    \bottomrule
    \end{tabular}
    \caption{Priors adopted for the BH primary-mass distributions associated with the three progenitor-metallicity models. The priors are chosen to preserve the overall structure of each fiducial metallicity model while allowing moderate variation in the location, width, and support of the mass distribution. Mass parameters are given in units of ${\rm M}_\odot$.}
    \label{tab:BH_priors}
\end{table*}

\bibliography{bib}
\bibliographystyle{aasjournal}

\end{document}